\documentclass{pnastwo}
\usepackage{pnastwof}
\usepackage{amssymb}
\usepackage{amsmath}
\usepackage[pdftex]{graphicx}
\usepackage{dcolumn}
\usepackage{bm}
\usepackage[round, sort&compress]{natbib}

\newcommand{\EQ}[1]{Eq.~(\ref{eq:#1})}

\newcommand{\FIG}[1]{Fig.~\ref{fig:#1}}

\newcommand{\fit}{F}		
\newcommand{\afit}{A}		
\newcommand{\efit}{E}		
\newcommand{\fitcoeff}{f}		
\newcommand{\mfit}{\bar{\fit}}	
\newcommand{\mafit}{\bar{\afit}}	
\newcommand{\mefit}{\bar{\efit}}	
\newcommand{\varfit}{\sigma}		
\newcommand{\afc}{f}
\newcommand{\gt}{g}		
\newcommand{\locus}{s}
\newcommand{\DOS}{\rho}
\newcommand{\adis}{\vartheta}
\newcommand{\edis}{\omega}
\newcommand{\EA}{S_A}		

\newcommand{\gene}[1]{\emph{#1}}
\newcommand{\beast}[1]{\emph{#1}}

\begin{document}
\title{%
Competition between recombination and epistasis can cause a transition from allele to genotype selection}
\author{Richard A.~Neher and Boris I.~Shraiman%
\affil{1}{Kavli Institute for Theoretical Physics, University of California, 
Santa Barbara,  CA 93106, USA}}
\maketitle
\begin{article}
\begin{abstract}
Biochemical and regulatory interactions central to biological networks are expected to cause extensive genetic interactions or epistasis affecting the heritability of complex traits and the distribution of genotypes in populations. However, the inference of epistasis from the observed phenotype-genotype correlation is impeded by statistical difficulties, while the theoretical understanding of the effects of epistasis remains limited, in turn limiting our ability to interpret data. Of particular interest is the biologically relevant situation of numerous interacting genetic loci with small individual contributions to fitness. Here, we present a computational model of selection dynamics involving many epistatic loci in a recombining population. We demonstrate that a large number of polymorphic interacting loci can, despite frequent recombination, exhibit cooperative behavior that locks alleles into favorable genotypes leading to a population consisting of a set of competing clones. When the recombination rate exceeds a certain critical value that depends on the strength of epistasis, this "genotype selection" regime disappears in an abrupt transition, giving way to "allele selection"-the regime where different loci are only weakly correlated as expected in sexually reproducing populations. We show that large populations attain highest fitness at a recombination rate just below critical. Clustering of interacting sets of genes on a chromosome leads to the emergence of an intermediate regime, where blocks of cooperating alleles lock into genetic modules. These haplotype blocks disappear in a second transition to pure allele selection. Our results demonstrate that the collective effect of many weak epistatic interactions can have dramatic effects on the population structure.
\end{abstract}
\dropcap{S}election acting on genetic polymorphisms in populations is a major force of evolution
\cite{McDonald_Nature_1991, Begun_PlosBiology_2007, Gerrish_Genetica_1998, Desai_Genetics_2007} and it is 
possible to identify specific loci under positive selection (e.g. the \gene{Adh} locus 
in  \beast{Drosophila} \cite{McDonald_Nature_1991}). Yet, the attribution of 
fitness differentials to specific allelic variants and combinations remains a great challenge \cite{Mackay_NatRevGen_2001}.  
Efforts to correlate quantitative phenotypes with genetic polymorphisms typically identify a small 
number of loci with a significant contribution to the observed phenotypic variance, but leave much
of the variance unaccounted for \cite{Barton_NatRevGen_2002}. This unaccounted variance is believed to arise 
from a large number of loci with small individual contributions, or be due  to epistasis and 
quite likely involves both effects. New studies  accumulate evidence that 
epistasis is widespread and accounts for a significant fraction of phenotypic variation (e.g. in  
yeast \cite{Brem_Nature_2005, Segre_NatureGenetics_2005, Schuldiner_Cell_2005}). 
Additional evidence for epistasis comes from crosses of mildly diverged strains, where the recombinant
progeny often has reduced average fitness, i.e.~ display outbreeding depression.
The reduction in fitness is attributed to the breakdown of favorable combination of alleles in the 
ancestral strains \cite{Dobzhansky_Genetics_1950}. Outbreeding depression is often observed in 
partly selfing organisms such as \beast{C.~elegans} \cite{Dolgin_Evolution_2007} or plants \cite{Parker_Evolution_1992},
species with strong geographic isolation such copepod \cite{Edmands_JHered_2008} or facultatively mating
organisms such as yeast \cite{Kuehne_CurrentBiology_2007}. While most recombinant genotypes are less fit, 
novel genotypes that perform better than either parental strain can be generated as well \cite{Wright_Genetics_1931}. 
Such outcrossing events could play an important role in evolution.

Competition between epistatic selection and recombination, explicit in the outbreeding depression phenomenon, is the focus of the present study. In the presence of epistasis, selection, by increasing the  frequency of favorable genotypes, establishes correlations between
alleles at different loci. Recombination on the other hand reshuffles alleles and randomizes  genotypes breaking up coadapted loci. Because recombination rate between any two loci is largely determined by their physical distance on the 
chromosome, the effect of genetic interactions depends on gene location.
It is known that functionally related 
genes tend to cluster \cite{Roy_Nature_2002, Hurst_NatRevGen_2004}, suggesting selection on gene order. 
Furthermore, chromosomes have regions of infrequent recombination, interspersed with recombination hotspots \cite{HapMap}. Does selection have a hand in defining low recombination regions?
To understand how evolution shaped genomes as we observe them today, we have to tackle the problem of 
how selection acts on many interacting polymorphisms for a large range of recombination rates \cite{Slatkin_NatRevGen_2008}.

\begin{figure}
\centering
\includegraphics[width=8.7cm]{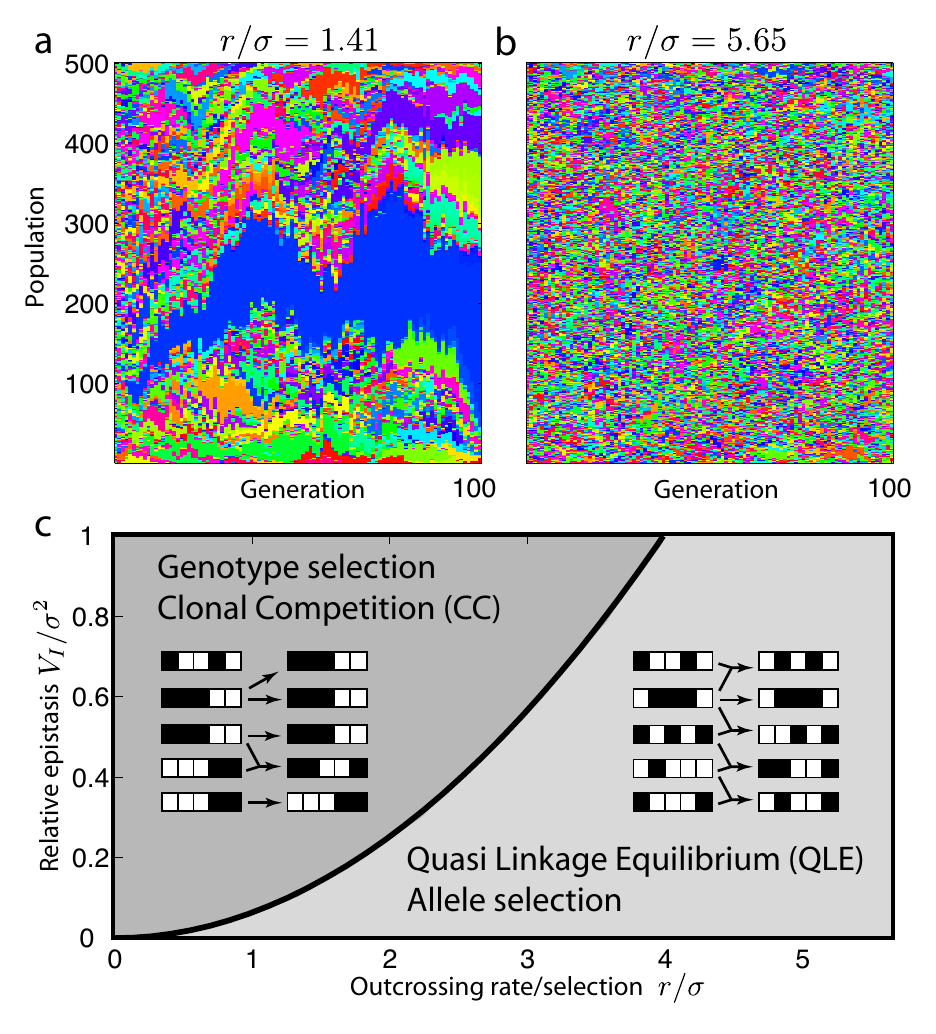}
\caption{\label{fig:phase_diagram}The two regimes of sexual reproduction. Panels \textbf{a} \& \textbf{b} show the
simulated time course of the genotype distribution in a population  of 500 individuals with epistatic fitness variance $V_I=\varfit^2=0.005$ and the outcrossing rate
$r=0.1$ (\textbf{a}) and $r=0.4$ (\textbf{b}). Like genotypes are assigned the same color and stacked on top of each other.
Sketches illustrating the population dynamics in the two cases are shown as insets in panel \textbf{c}.
At low outcrossing rates, fit genotypes can proliferate. The genotype distribution rapidly coarsens and clones form (horizontal stripes in panel \textbf{a}). 
With frequent outcrossing, genes are rapidly reshuffled and genotypes do not persist over many generations, resulting
in the pointillist pattern in panel \textbf{b}. Fixation happens at later time and is not shown.
Panel \textbf{c}: The two regimes are separated by a sharp boundary set by the strength of epistasis. 
For $r<r_c$, the population dynamics is described by clonal competition (CC); for $r>r_c$ by 
quasi linkage equilibrium (QLE). }
\end{figure}

Standing variation harbored in natural population provides important raw material for selection to act upon, 
in particular after a sudden change in environments or hybridization events \cite{Teotonio_NatGen_2009}. In 
such a situation, selection will reduce genetic variation until a new mutation-selection equilibrium is reached. 
Here, we show that the selection dynamics on standing variation at a large number of loci can be strongly affected by epistasis, 
even if the individual contribution of each locus is small. The competition between
selection on epistasis and  recombination gives rise to two distinct regimes at high and low recombination rates
separated by a sharp transition. The population dynamics in the 
two regimes is illustrated in \FIG{phase_diagram}a,b:  i) the ``clonal competition" (CC)  regime which occurs for recombination rates $r<r_c$ and  ii) the Quasi Linkage Equilibrium (QLE) regime for $r>r_c$. 
The different nature of the two regimes is best understood by considering the limiting cases of no and frequent
recombination. 
In the case of purely asexual reproduction, selection operates on entire genotypes and results in clonal expansion of the fitter ones. The genetic variation present in the initial population
is lost on a timescale inversely proportional to the average magnitude of fitness differentials between genotypes present in the population. Successful genotypes persist in time, which is apparent as continuous broad stripes of one 
color in \FIG{phase_diagram}a. The amplification of a small number of fit genotypes induces strong 
correlations or linkage disequilibrium among loci. In presence of epistasis, a little recombination does not 
change this picture qualitatively, as most recombinant genotypes are less fit than the prevailing clones and 
novel successful clones are rare. Nevertheless recombination is very important because it continuously introduces new genotypes leading to an increase in fitness attained by the population at long times.
In the limit of high recombination genotypes are short-lived and essentially unique, resulting in a ``pointillist'' color 
pattern in \FIG{phase_diagram}b. Each allelic variant is therefore selected on the basis of its effect on fitness, 
averaged over many possible genetic backgrounds. The time scale on which allele frequencies change is 
given by the inverse of these marginal fitness effects. The term "linkage equilibrium" in QLE refers to the negligible
correlations between loci, which are constantly reshuffled by recombination. 

As we show below, the transition between the two regimes sharpens as the number of segregating loci 
$L$ increases. The sharpening of the transition is related to the different scaling of the time scale of selection in the two regimes. 
For large $L$, the marginal fitness effects of individual loci become small compared to fitness differentials among
individuals (assuming they are all of similar size, this ratio decreases as
$\sim 1/\sqrt{L}$). Hence, the dynamics in the QLE regime slows down compared to the CC regime
as $L$ increaes. The CC and QLE regimes correspond
to different regions of the parameters space spanned by the relative strength of epistasis and the ratio of 
outcrossing or recombination rate to the strength of selection, as sketched in \FIG{phase_diagram}c. The QLE 
dynamics was first described by Kimura \cite{Kimura_Genetics_1965} in the limit of weak selection/fast 
recombination for a pair of bi-allelic loci and subsequently generalized to multi-loci systems 
\cite{Barton_Genetics_1991, Nagylaki_Genetics_1993}. 
The possibility  of a collective behavior  involving linkage disequilibrium on many loci and selection 
effectively acting on the whole chromosome as a unit has been pointed out before in the context of 
overdominance by Franklin and Lewontin \cite{Franklin_Genetics_1970} in the strong selection limit.  
However, these studies of the two different limits do not reveal the breakdown of QLE and the transition 
to CC as the generic behavior of  multi-locus epistatic systems.

To underscore the general nature of the results, we shall consider two different models of epistasis. 
The first model will follow the common treatment of epistasis in quantitative traits which 
assumes that the epistatic contribution to fitness is disrupted when the parental genes are 
mixed in sexual reproduction \cite{Lynch_1998, Falconer_1996}. 
This assumption becomes exact when the epistatic component of fitness of a specific genotype 
is a random number (which depends on the genotype, but is fixed in time) and we shall call this model the random epistasis (RE) model. Within the 
RE model, any change in the genotype randomizes the epistatic component of fitness so that 
the latter is not heritable when non-identical parents mate. It is, however, faithfully passed on 
to the offspring in asexual reproduction. For the RE model, genomes are propagated asexually 
with probability $1-r$ and with probability $r$ are a product of mating where 
all genes are reassorted, as would be exactly correct if all genes were on different chromosomes. 
This model of facultative mating approximates reproductive strategies common in fungi (e.g. yeast) or nematodes and plants.  
As a more realistic alternative, we shall also study a model with only pairwise interactions between loci \cite{Hansen_TheoPopBiol_2001}. This pairwise epistasis (PE) model allows epistatic contribution to be partly heritable, as interacting pairs have a chance to be inherited together \cite{Bulmer_1980}. 
For the PE model, we assume that all genes are arranged on one chromosome with a uniform crossover
rate $\rho$, which allows us to explore haplotype block formation and implications for recombination
rate evolution. 

The strength of selection is determined by the variance $\varfit^2$ of the distribution of fitness in the 
population. Within our models, the fitness $\fit(\gt)$ of a genotype $\gt$ is the sum of an additive 
component $\afit(\gt)$ representing independent contributions of alleles and an epistatic part $\efit(\gt)$. 
For the RE model, the latter is a random number drawn from Gaussian distribution, while  for the PE model it is a 
sum of pairwise interactions with random coefficients $\fitcoeff_{ij}$. The variances 
$V_A$ and $V_I$ of the distributions of $\afit(\gt)$ and $\efit(\gt)$ add up to $\varfit^2$ and their relative
magnitude determines the importance of additive effects compared to epistasis.  The two different models and their parameters
are given explicitly in the methods section. For the sake of simplicity, we assume haploid genomes. 
Random and pairwise epistasis represent two opposite extremes in the complexity of epistasis.
While the pairwise model is more realistic, the generic behavior is most clearly demonstrated using the RE model
with random gene reassortment and facultative mating.

\section{Results}
\begin{figure}
\centering
\includegraphics[width=8.7cm]{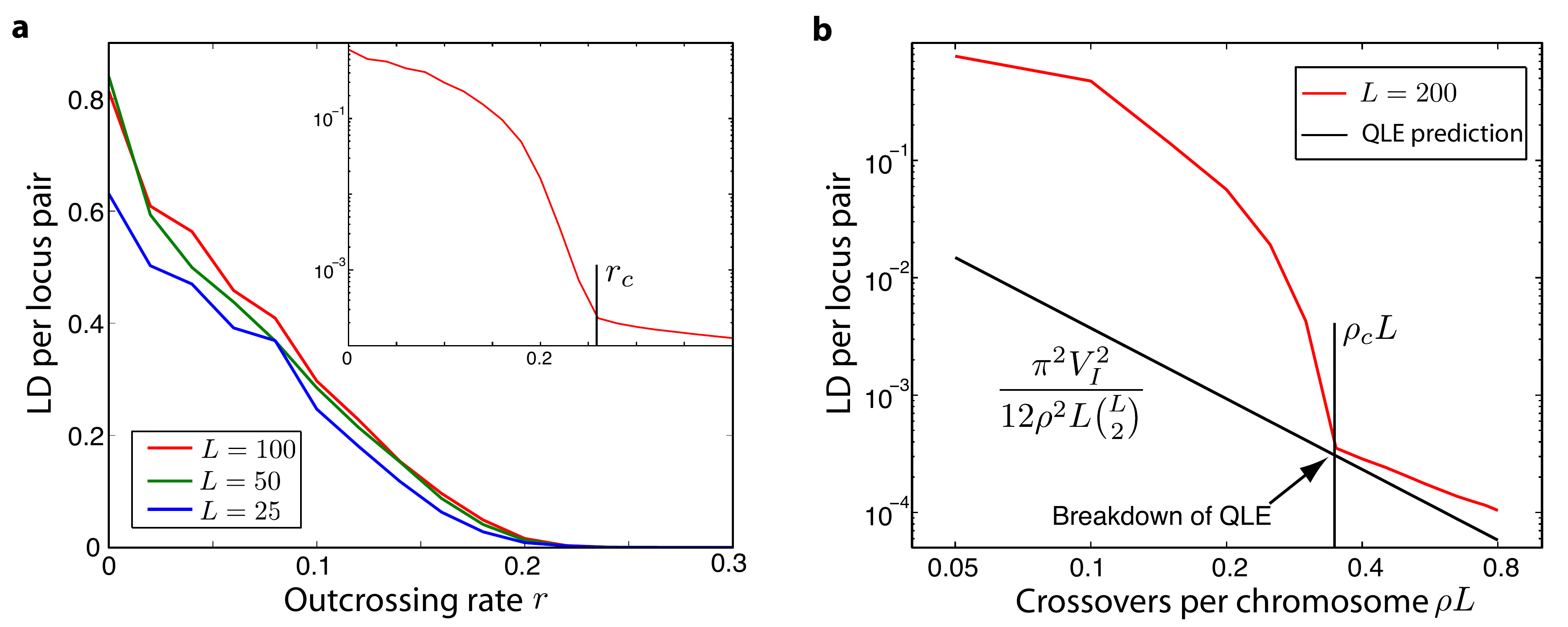}
\caption{\label{fig:LD} The clonal competition regime is characterized by extensive linkage disequilibrium. 
\textbf{a} Random epistasis model: For small $r$, the LD per locus pair is of order one
and fairly independent of $L$. The inset shows the data for $L=100$ on a logarithmic scale and a mark at the value of $r_c$.
The LD for $r>r_c$ is due to sampling noise, see Figure S1. 
\textbf{b} Pairwise epistasis model: 
For pairwise epistasis, the QLE approximation gives explicit predictions for LD, which describes the 
observed LD very accurately for $\rho>\rho_c$, black line. For $\rho<\rho_c$, LD is a much larger than the QLE prediction. 
For both panels, LD is measured when allelic entropy has decayed 30\% from the initial value ($\varfit^2=0.005$, $V_A=0.1\varfit^2$ and $V_I=0.9\varfit^2$). 
In panel \textbf{a}, $N=10^5$ and the data shown is averaged over 100 realizations. 
To avoid boundary and finite size effects, we 
used $N=10^6$ and a circular chromosome for panel \textbf{b} and averaged over 10 realizations.
}
\end{figure}

\paragraph{Two regimes of selection dynamics.}
We performed extensive computer simulation of our two models for
different relative strength of epistasis, $L=25-200$ loci and  populations sizes between $N=500$ and $10^6$.  
We initialize simulations in a genetically diverse state as would result from multiple crossings of two diverged strains and
examine the evolution under selection and recombination. The two regimes differ strongly in
the amount of linkage disequilibrium  (LD) (see Methods) build up by selection. 
Panel \textbf{a} of \FIG{LD} shows the average LD per locus pair for the RE model as a function of the outcrossing rate $r$.
For $r<r_c$, the LD per locus pair is of order one and independent of $L$ or $N$, indicating genome-wide LD. 
LD builds up despite a large number of  different genotypes in the population interbreeding constantly. 
For $r>r_c$, the LD is much smaller, with the observed value determined by the
sampling noise due to the finite population size (see inset of Fig 2a and supplementary Figure S1). 
Similar behavior occurs in the PE model, as shown 
in panel \textbf{b}. Above a critical recombination rate $\rho_c$, the observed linkage disequilibrium is time independent and well described by the QLE approximation \cite{Kimura_Genetics_1965, Barton_Genetics_1991} 
(straight line, see supplement). 
The QLE approximation (in the high $\rho /\sigma$ limit) 
predicts LD to be proportional to the strength of {\it pairwise} epistasis
Below $\rho_c$, the observed LD is dramatically larger than the QLE expectation. 
Here, recombination is sufficiently infrequent such that genotypes with a 
synergistic alleles are amplified faster than they are taken apart by recombination, see below. 
As a result, the few fittest genotypes grow exponentially in number, leading to the strong 
correlation in the occurrence of cooperating alleles, independent of physical linkage 
(i.e. proximity on the chromosome). This extensive LD leads to a complete failure when extrapolating 
results valid in the high recombination regime across the transition.
The relevant quantity that determines whether fit genotypes can be maintained
is the probability that no crossover occurs, which is given by $e^{-\rho L}$. 
Hence, $\rho_c$ is inversely proportional to $L$.

\begin{figure}	
\centering
\includegraphics[width=8.7cm]{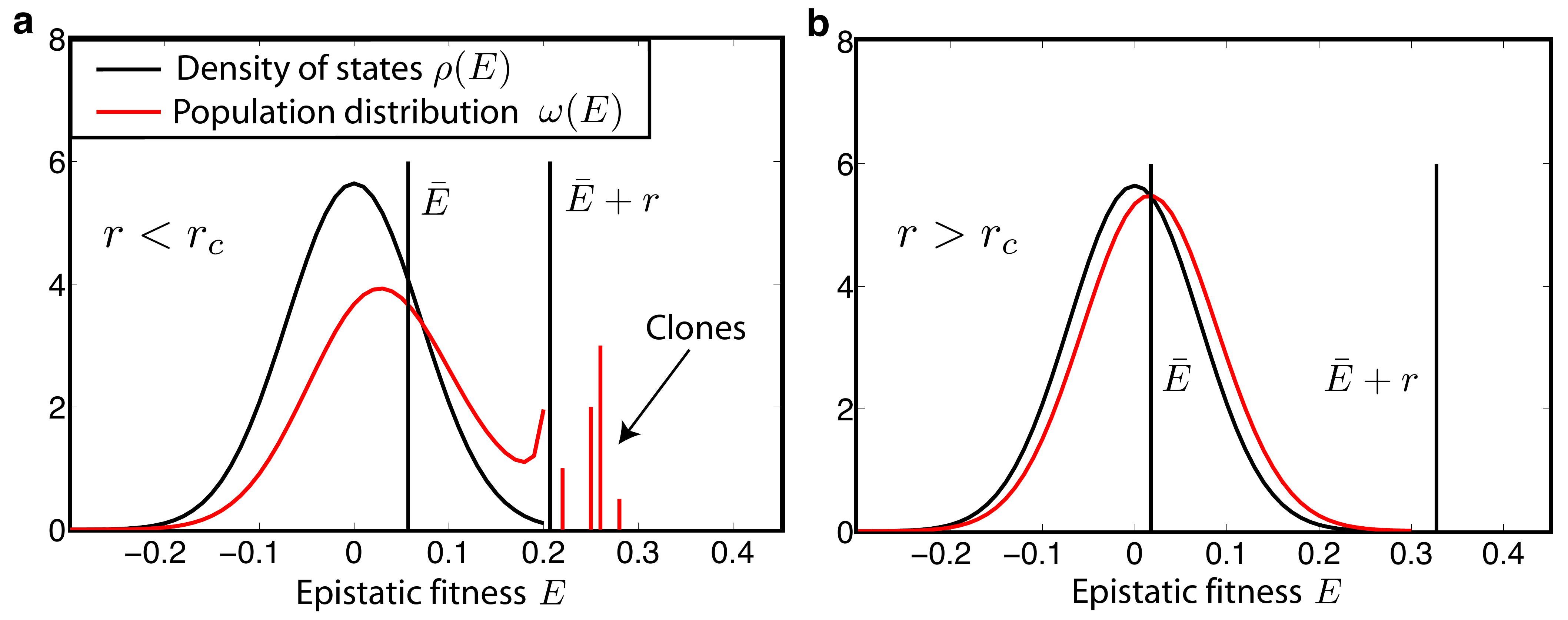}
\caption{\label{fig:MFT_illustration}The break-down of QLE. Panel \textbf{a}: 
When the recombination rate decreases below $r_c$, some individuals have epistatic fitness $E$ 
larger than $\bar{E}+r$, and the QLE solution for the distribution of epistatic fitness 
in the population breaks down. Individuals to the right of $\bar{E}+r$ form clones that grow exponentially and
the population condenses into a small number of genotypes. 
Panel \textbf{b}: For $r>r_c$, even the largest epistatic fitness contributions do not result in a 
growth advantage that exceeds the recombination rate. 
 }
\end{figure}

\paragraph{Self-consistency condition for QLE.}
The fitness of a genotype can be decomposed as $\fit=\afit+\efit$,
where $\afit$ is the heritable additive part and $\efit$ is the non-heritable
epistatic part. As a coarse-grained descriptor of the population, we consider the
joint distribution $P(\afit, \efit; t)$ of the fitness components. In the QLE
state, $P(\afit, \efit; t)$ evolves approximately as
\begin{equation}
\label{eq:meanfield}
\partial_t P(\afit, \efit; t)= (\fit-\mfit -r )P(\afit, \efit; t)+r\DOS(\efit) \adis(\afit; t)
\end{equation}
The first term accounts for the exponential growth of genotypes with fitness advantage
$\fit-\mfit$ and the loss due to recombination at rate $r$.
The second term accounts for the production of genotypes through recombination. 
To a good approximation, the distribution of $\afit$ among recombinant
offspring is identical to that among the parents $\adis(\afit)=\int d\efit\; P(\afit, \efit)$, which in turn is approximately 
Gaussian \cite{Turelli_Genetics_1994}. The distribution of $\efit$ among recombinant
offspring is independent of the parents and a random sample from the distribution of epistatic 
fitness $\DOS(\efit)$, which in our models is a zero-centered Gaussian. The latter is exactly true for the RE model
and holds approximately for the PE model, where the correlation of
$\efit$ between ancestor and offspring halves every generation \cite{Bulmer_1980}. \EQ{meanfield}
admits the factorized solution $P(\afit, \efit; t)=\adis(\afit; t)\edis (\efit)$ with 
$\partial_t \adis(\afit;t)=(\afit-\mafit)\adis(\afit;t)$ 
and a time-independent distribution of $\efit$
\begin{equation}
\label{eq:box_PE}
\edis(\efit)= \frac{r\DOS(\efit)}{r+\mefit-\efit},
\end{equation}
where $\mefit$ is determined by the condition that $\edis(\efit)$ has to be normalized.
This solution exists only if $\efit<r+\mefit$ for all genotypes; otherwise, fit
genotypes escape recombination and grow as clones. 
These two scenarios are illustrated in \FIG{MFT_illustration}.

The normalization condition can be fulfilled only if $r$ is larger than some $r_c$\footnote{ 
$\DOS(\efit)$ has to go to zero faster than linear for $r_c$ to exist.}. The value of $r_c$ is
proportional to the maximal $\efit$ and hence proportional to the strength of
epistasis $\sqrt{V_I}$. However, it is not the absolute maximum of $\efit$ 
among all possible $2^L$ genotypes that determines $r_c$, but the maximal $\efit$ that is
encountered by the population before fixation. Hence $r_c$ depends on the population size
and the functional form of this dependence is determined by the upper tail of the distribution
 $\DOS(E)$. For the Gaussian distribution used here, $r_c\sim \sqrt{\ln (rN\tau)}$, 
where $\tau$ is the time scale of QLE dynamics discussed below.. The product 
$rN\tau$ then is the number of genotypes generated through recombination before fixation.
A more detailed discussion is given in the Supplementary information.

The breakdown of the QLE state has some similarity to the error-threshold transition of a
quasi-species model \cite{Eigen_Naturwissenschaften_1971} in a rugged fitness
landscape \cite{Franz_JPA_1997}: Recombination of epistatic loci acts as deleterious 
mutations and prevents the emergence of quasi-species or
clones \cite{Boerlijst_PRSB_1996, Park_PRL_2007} for $r>r_c$. 

\begin{figure}
\centering
\includegraphics[width=8.7cm]{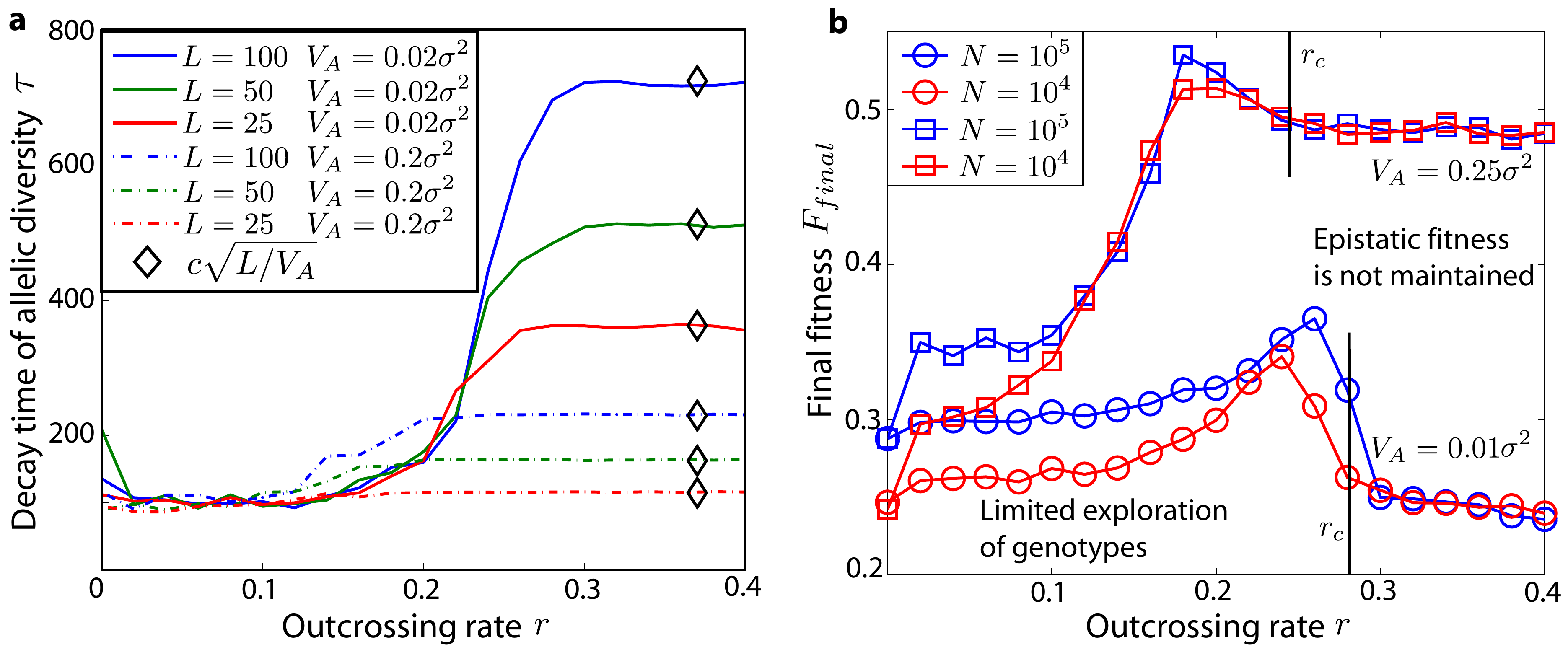}
\caption{\label{fig:decay_peak} Panel  \textbf{a} shows the time $\tau$ it takes to reduce 
the allelic entropy by 30\% as a function of $r$ for different parameters. For small $r$, $\tau$ is independent of $L$
but increases with $r$. For $r>r_c$, $\tau$ settles at $c \sqrt{L/V_A}$ (black diamonds) in accord with the theoretical
prediction for single locus dynamics (with {\it c} a fitting parameter). 
Additional data for $V_A=0$, $V_A=0.5\varfit^2$ and a collapse 
confirming the scaling of $\tau$ is shown in supplementary figure S2. 
Panel \textbf{b}: The fitness of the fixated genotype $\fit_{\rm{final}}$ as a function of $r$ for two different strength of epistasis.
At $r=0$, the final fitness only depends on the population size $N$ and is independent of the strength of epistasis. 
$\fit_{\rm{final}}$ increases with $r$, followed by a pronounced drop right below $r_c$. 
Above $r_c$, $\fit_{\rm{final}}$ is almost constant and is independent of $N$.
In both panels, $\varfit^2=0.005$. Data is averaged over 25 realizations in panel \textbf{a} and over 100 realizations in panel \textbf{b}. $L=100$ in panel \textbf{b}.
}
\end{figure}

\paragraph{Maintenance of genetic diversity.}
The transition between the two regimes leaves its imprint in virtually every quantity of interest in population genetics. 
For instance, the characteristic time for the decay of genetic diversity, $\tau$, (which we quantify via allele entropy, see Methods)
scales differently with $L$ in the two regimes, as shown in \FIG{decay_peak}. 
At low outcrossing rates, $\tau$ depends only on 
the total variance in fitness and neither on the number of loci nor the relative strength of additive contributions. This is 
consistent with the notion that in the CC regime genotypes are the units on which selection acts. With more frequent outcrossing,
$\tau$ tends to be larger for weak additive contributions and large $L$. Beyond a certain outcrossing rate $r_c$,
$\tau$ becomes independent of $r$ attaining a value inversely proportional to the additive contribution of the individual loci independent of $V_I$ (black diamonds in Fig 3a). This observation confirms our  assertion that for $r>r_c$, outcrossing 
decouples the loci and that the allele frequencies evolve independently under the action of the additive component of fitness.
Given an additive variance $V_A$, the typical single locus fitness differential is $\afc\sim \sqrt{V_A/L}$ such that $\tau$ grows
as $\sqrt{L}$ for $r>r_c$. 
To uncover the universal behavior in the vicinity of the transition in the limit of large genomes, we show that the 
data for different $V_I$, $V_A$ and $L$ collapses onto a single master curve after appropriate rescaling of the axis, see Fig.~S2.
This scaling collapse demonstrates the existence of a sharp transition in the limit $L\to \infty$, the scaling of $\tau$ with 
$\sqrt{L}$ and shows that $r_c$ is proportional to $\sqrt{V_I}$, as expected from the self-consistency argument
outlined above and sketched in \FIG{phase_diagram}c.
The suppression of allele dynamics by $1/\sqrt{L}$ in the QLE regime is at the basis of Fisher's infinitesimal model
put forward to explain sustained response to selection \cite{Barton_NatRevGen_2002}. 
In one generation, the allele frequencies change by approximately $\afc$, which can 
be sustained over $\sim \afc^{-1}$ generations. The mean fitness increases by $V_A$ per generation,
consistent with Fisher's theorem \cite{Lynch_1998, Nagylaki_Genetics_1993}. Our results show,
that epistasis causes the breakdown of the infinitesimal model for $r<r_c$. 
The pairwise epistasis model is more complex than the random epistasis model, since the partition of the fitness variance 
in additive and epistatic contribution depends on the allele frequencies and epistasis 
is ``converted'' into additive fitness as the population approaches fixation \cite{Turelli_Evolution_2006} (a detailed account will be published elsewhere).

The properties of the genotype which will eventually fixate in the population
depends on the regime in which it was obtained. 
We find, that the fitness of this fixated genotype depends non-monotonically on the outcrossing rate and 
peaks just below the transition, see \FIG{decay_peak}. This  can be understood as follows. 
Without recombination, the final state can be no fitter than the fittest genotype initially present. 
With some recombination, the population explores a greater number genotypes, potentially finding ones with higher fitness
so that the fitness of final state increases with $r$ in the CC regime. A similar benefit of infrequent recombination due to 
exploration of genotype space has been studied in the context of virus evolution for additive fitness functions
 \cite{Rouzine_Genetics_2005}. 
As  genotype selection gives way to allele selection, different loci decouple and the
epistatic contribution to fitness is missed, leading to possible fixation of less fit genotypes and 
a sharp drop of the final fitness  $r$ approaches $r_c$. 
The dependence of the final fitness on the population size $N$ highlight the distinct properties the dynamics in the 
two regimes: In the QLE regime, the final fitness is virtually identical for different $N$. This is a consequence of the
fact that the relevant dynamical variables are allele frequencies, which are well sampled by $\mathcal{O}(N)$ individuals.
Fluctuation of the allele frequencies are therefore negligible and the dynamics is essentially deterministic. This is different
in the CC regime, where the dynamics is driven by the generation of rare, exceptionally 
fit genotypes. The rate, at which genotypes are generated is proportional to the $N$, resulting in a pronounced
dependence on the population size. QLE ceases to be deterministic once the marginal fitness effects become 
comparable to inverse population size and random genetic drifts overwhelms selection, see Fig.~S3 in the Supplementary
information.

\begin{figure}
\centering
\includegraphics[width=8.7cm]{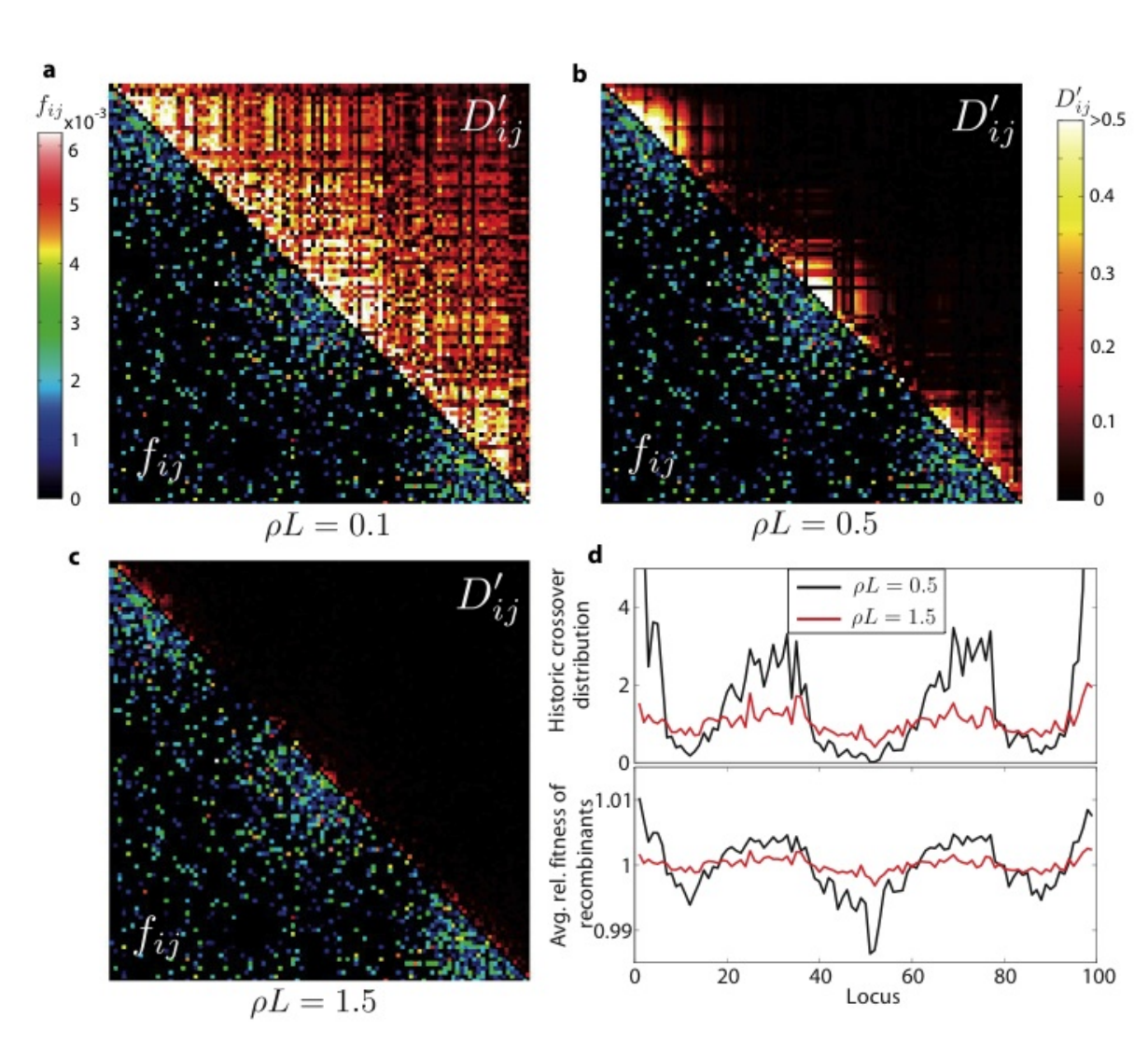}
\caption{\label{fig:Blocks}
Clonal competition, modular selection and quasi linkage equilibrium.
Above the diagonal, panels \textbf{a}, \textbf{b} and \textbf{c} show the LD measured as $D_{ij}'$ between two 
loci $i$ and $j$ for a linear chromosome of length $L=100$ at three different crossover rates $\rho$. 
Below the diagonal, the interaction matrix $f_{ij}$ is shown (the same in all three panels). 
At low $\rho$, the sparse long range
interactions suffice to produce genome wide LD. At intermediate $\rho$, distant part of the genome are decoupled,
but the more strongly interacting clusters still show high LD, which vanishes at even higher recombination rates.
Panel \textbf{d}, top: the distribution of historic crossovers. 
Bottom: The relative fitness of recombinants as a function
of the crossover location. LD was measured when allelic entropy was at 90\% of the initial value, $\varfit^2=V_I=0.005$ and $N=10^6$.
}
\end{figure}

\paragraph{Selection on genetic modules.}
So far, we assumed that each pair of loci is equally likely to interact epistatically, regardless of their physical 
distance on the chromosome. However, there is evidence that the order of genes along the chromosome is
far from random and that related genes tend to cluster \cite{Roy_Nature_2002, Hurst_NatRevGen_2004}. 
To emulate such a situation we use the PE model and construct an interaction matrix $\fitcoeff_{ij}$ where arbitrary pairs
interact with a small probability while clusters of neighboring genes interact with a high probability (see Methods).
For such a hierarchical epistatic structure, we observe, as a function of increasing crossover rate $\rho$, a sequence of two transitions which define, sandwiched between CC and QLE, an intermediate Modular Selection (MS) regime, where the genome-wide LD characteristic of the CC regime has broken down to a set of modular blocks which are in quasi linkage equilibrium with each other. The resulting linkage disequilibrium patterns are shown in \FIG{Blocks}. The observed
block structure of LD in the MS regime resembles haplotype blocks \cite{HapMap, Slatkin_NatRevGen_2008}, which are normally associated with 
regions of little recombination flanked by recombination hotspots. Indeed, the  cumulative recombination history 
of the chromosomes in the population show a very heterogenous recombination distribution, as shown in panel \textbf{d} of \FIG{Blocks}. Yet, here the origin of these blocks is not intrinsically low recombination (i.e. physical linkage) but the collective effect of epistatic selection: the surviving individuals have recombined more often in regions of low epistasis than in regions of high epistasis, 
even though the attempted crossovers are uniformly distributed along the chromosome.
Clusters of epistatic interaction can therefore exert selective pressure to lower recombination within the cluster. 
This lack of recombinant survival has been observed in experiments with mice \cite{Petkov_PLOSgenetics_2005}, 
where inbreeding results in strong selective pressure on localized clusters of genes generating blocks with 
high LD and reduced effective recombination.

\section{Conclusion}
To summarize, we have shown that the competition of epistatic selection and recombination can give rise to distinct regimes of population dynamics, separated by a 
transition that becomes sharp for large number of interacting loci. 
The QLE and CC regimes are realizations of the opposing views on evolution of R.A. Fisher and 
S. Wright. For $r>r_c$ alleles are selected for the their additive contributions while selection acts on whole genotypes
for $r<r_c$.
The fundamental differences between these two regimes show up most clearly in the different scaling properties 
of the total LD and the decay time of genetic diversity. 
In the low recombination regime, LD is produced independent of physical linkage by the collective
effect of many interactions. In the high recombination regime, LD can be attributed to specific interactions between
pairs of loci and its value, determined by the ratio of the interaction strength and the rate of recombination between the loci, is small. 
Our results not only apply to the transition between genotype and allele selection, but also to localized clusters
of interacting genes on the chromosome. Whenever the epistatic fitness difference between different allelic compositions
of a cluster exceeds the recombination rate of the cluster, the fittest will amplify exponentially. Since such clusters are often small
\cite{Petkov_PLOSgenetics_2005} ($\sim$ Mb) their recombination rates are low (cM or less) - hence fitness 
differentials below one percent can suffice to establish CC dynamics. Selective pressure to reduce recombination load, i.e.
the fitness loss through recombination, will therefore favor the evolution of clusters of interacting genes and might  
be an important driving force for the evolution of recombination rate \cite{Nei_Genetics_1967,Barton_Genetics_2005}.  The effects described above may provide an explanation for the functional clustering associated with low and high LD regions reported in HapMap \cite{HapMap}.

\begin{acknowledgments}
We would like to thank Michael Elowitz and Marie-Anne Felix for comments on the manuscript and acknowledge financial support from NSF grant  PHY05-51164.
\end{acknowledgments}

\section*{Methods}
\paragraph{Random epistasis model.}
 A genotype $\gt$ is described
by $L$ binary variables $\locus_i=\pm 1$, $i=1,\ldots, L$. To each genotype
we assign a fitness
\begin{equation}
\fit(\gt)=\afc\sum_i^L \locus_i+\xi(\gt).
\end{equation}
The first term is the sum of the additive fitness contributions of the individual loci, each of which has equal
magnitude $\afc=\sqrt{V_A/L}$. The second term is the non-heritable epistatic
fitness, where $\xi(\gt)$ is drawn from a normal distribution with zero mean and variance $V_I$. 
For a uniform distribution of genotypes, the additive fitness variance is $V_A$, 
the epistatic variance is $V_I$, and the total variance is $\varfit^2=V_A+V_I$. 
\paragraph{Pairwise epistasis model.}
Here, we consider epistasis due to pairwise interactions between the different loci. Such pairwise interactions
correspond to $\locus_i\locus_j$ terms in the fitness function. 
The fitness of a particular genotype $\gt$ is determined by the independent effects of the
individual loci and the sum of the interactions between all pairs. 
\begin{equation}
\fit(\gt)=\afc\sum_i^L \locus_i+\sum_{i<j}\fitcoeff_{ij}\locus_i\locus_j.
\end{equation}
When assuming uniform epistasis between all possible pairs, we draw the interaction strength $\fitcoeff_{ij}$ 
from a Gaussian distribution with zero mean and variance $\frac{2V_I}{L(L-1)}$.
\paragraph{Clustered epistasis.}
To mimic localized clusters of strongly interacting genes on a weakly interacting background, we constructed
the matrix of $\fitcoeff_{ij}$'s as follows. The sparse background epistasis was modeled by assigning each 
$\fitcoeff_{ij}$ a Gaussian random number with probability $p=0.1$ and zero otherwise. 
Then we built three epistatic clusters with centers $c_k= 10, 50, 90$ by adding  
a Gaussian random number to each $\fitcoeff_{ij}$ with probability 
$p=\exp\left(\frac{(i-c_k)^2+(j-c_k)^2}{2r^2}\right)$ with $r=10$ for $k=1, 2, 3$. All $\fitcoeff_{ij}$ were rescaled
such that $\sum_{i<j}f_{ij}^2=V_I$.

\paragraph{Selection.}
Our model assumes non-overlapping generations. In each generation a pool of 
gametes is produced, to which each individual contributes a number of copies of
its genome which is drawn from a Poisson distribution with parameter $\exp(\fit(\gt)-\mfit)$.
\paragraph{Gene re-assortment.}
To model gene re-assortment in a facultatively mating population, two gametes are chosen
with probability $r$ and a new genotype is formed by assigning each locus 
the allele of one or the other parent at random. Otherwise, the new genotype is an exact copy of one gamete.
\paragraph{Crossovers.}
Given a crossover rate $\rho$ per locus, the number of crossovers is drawn from a Poisson distribution 
with parameter $(L-1)\rho$ and the crossover locations are chosen at random. When the number of 
crossovers is zero, the offspring inherits the entire genome from one parent. To model circular chromosomes, 
the number of crossovers is multiplied by two enforcing an even number of crossovers.

\paragraph{Measuring genetic diversity.}
The allele entropy is a convenient descriptor of genetic diversity that is readily calculated
from the evolving population. It is defined as $\EA=-\sum_i \left[\nu_i\ln \nu_i+(1-\nu_i)\ln (1-\nu_i)\right]$, where $\nu_i$ 
is the allele frequency at locus $i$. 

\paragraph{Measuring linkage disequilibrium.}
LD is the deviation of the frequency of a pair of alleles from the random expectation on the basis of the individual 
allele frequencies, i.e. $D_{ij}=\langle \locus_i\locus_j\rangle-\langle \locus_i\rangle\langle\locus_j\rangle$.
Kimura showed \cite{Kimura_Genetics_1965} that in QLE $\psi_{ij}=\frac{D_{ij}}{\nu_i\bar{\nu}_i\nu_j\bar{\nu_j}}$ is time independent despite changing allele frequencies $\nu_i$ and $\nu_j$ ($\bar{\nu}_i=1-\nu_i$). To measure
genome wide LD, we calculate the sum of all squared LD terms $\sum_{i<j}\psi_{ij}^2$. Pairs with $\nu_i$ or $\nu_j$ smaller than 0.01 or larger than 0.99 were omitted. 
A different normalization is used in \FIG{Blocks}, where 
$D_{ij}'=\frac{|D_{ij}|}{4\max(\min(\nu_i\nu_j,\bar{\nu_i}\bar{\nu_j}),\min(\nu_i\bar{\nu_j},\bar{\nu_i}\nu_j)}$ is shown, see Ref. ~\cite{Slatkin_NatRevGen_2008} for a recent review.

\end{article}







\end{document}